\documentstyle[epsf,eepic,11pt]{article}
\def\Fsquare(#1,#2){
\hbox{\vrule$\hskip-0.4pt\vcenter to #1{\normalbaselines
\hrule\vfil\hbox to #1{\hfill$#2$\hfill}\vfil\hrule}$\hskip-0.4pt
\vrule}}
\def\Addsquare(#1,#2){\hbox{$
        \dimen1=#1 \advance\dimen1 by -0.8pt
        \vcenter to #1{\hrule height0.4pt depth0.0pt%
        \hbox to #1{%
        \vbox to \dimen1{\vss%
        \hbox to \dimen1{\hss$~#2~$\hss}%
        \vss}%
        \vrule width0.4pt}%
        \hrule height0.4pt depth0.0pt}$}}

\newcommand{\NN}{\nonumber}

\newcommand{\hitotu}[1]{\hbox{$\hskip0.0em\raisebox{0.0ex}{\fbox{${#1}$}}$}}
\newcommand{\hutatu}[2]{\hbox{$\hskip0.0em\raisebox{1.38ex}{\fbox{${#1}$}}\raisebox{-1.38ex}{\hskip-1.12em\fbox{${#2}$}\hskip0.1em}$}}
\newcommand{\ket}[1]{|{#1}\rangle}
\newcommand{\IFT}{\int_{-\infty}^{\infty}dx e^{ikx}}
\newcommand{\tJ}{{\it t-J}\,}
\newcommand{\suka}{\left.}
\newcommand{\bou}[1]{\right|_{#1}}
\newcommand{\BEA}{\begin{eqnarray}}
\newcommand{\EEA}{\end{eqnarray}}
\newcommand{\BE}{\begin{equation}}
\newcommand{\EE}{\end{equation}}

\newcommand{\fT}{{\cal T}}
\newcommand{\fH}{{\cal H}}
\newcommand{\fh}{\pi}
\newcommand{\fZ}{{\cal Z}}
\newcommand{\fq}{{\tt q}}
\newcommand{\Tr}{{\rm Tr}\,}
\newcommand{\LR}{\left(}
\newcommand{\RR}{\right)}
\newcommand{\CMP}[3]{{\it Comm. Math. Phys} {\bf #1} {(#2)} {#3}}

\newcommand{\IJMP}[3]{{\it Int. J. Mod. Phys.} {\bf #1} {(#2)} {#3}}
\newcommand{\JP}[3]{{\it J. Phys.} {\bf #1} {(#2)} {#3}}

\newcommand{\JSPS}[3]{{\it J. Phys. Soc. Jpn.} {\bf #1} {(#2)} {#3}}
\newcommand{\MPL}[3]{{\it Mod. Phys. Lett.} {\bf #1} {(#2)} {#3}}
\newcommand{\NP}[3]{{\it Nucl. Phys.} {\bf #1} {(#2)} {#3}}

\newcommand{\PL}[3]{{\it Phys. Lett.} {\bf #1} {(#2)} {#3}}
\newcommand{\PR}[3]{{\it Phys. Rev.} {\bf #1} {(#2)} {#3}}
\newcommand{\PRL}[3]{{\it Phys. Rev. Lett.} {\bf #1} {(#2)} {#3}}
\newcommand{\PTP}[3]{{\it Prog. Theo. Phys.} {\bf #1} {(#2)} {#3}}
\newcommand{\Zeit}[3]{{\it Z. Phys.} {\bf #1} {(#2)} {#3}}
\begin{document}
\begin{center}
\vspace*{0.5cm}
{\large{\bf Anti-Symmetrically Fused Model and Non-Linear Integral 
            Equations in the Three-State Uimin-Sutherland Model}}  
\vskip 1.2cm
{\large{\bf Akira Fujii}}
\vskip 0.2cm
{\small {\sl KEK, Tanashi \\ Midori-cho 3-2-1 Tanashi \\
             188-8501 Tokyo, Japan}}             
\vskip 0.1cm
and
\vskip 0.1cm
{\large{\bf Andreas Kl{\"u}mper}}
\vskip 0.2cm
{\small{\sl Universit{\"a}t zu K{\"o}ln \\
            Institut f{\"u}r Theoretische Physik \\
            Z{\"u}lpicher Str.77 \\
            50937 K{\"o}ln, Germany}}
\end{center}
\vskip1cm
\begin{abstract}
We derive the non-linear integral equations determining the 
free energy of the three-state pure bosonic Uimin-Sutherland model. 
In order to find a complete set of auxiliary functions, the anti-symmetric 
fusion procedure is utilized. We solve the non-linear integral 
equations numerically and see that 
the low-temperature behavior coincides with that predicted
by conformal field theory. 
The magnetization and magnetic susceptibility are also calculated by means 
of the non-linear integral equation.
\end{abstract} 
\begin{flushleft}
\end{flushleft}
\baselineskip = 22pt 
\newpage
\section{Introduction}
The Bethe ansatz (BA) method \cite{bethe} 
is a most fundamental approach
to the study of one and two-dimensional exactly solvable lattice 
models. 
In the BA strategy, 
the quantum transfer matrix (QTM) method has been found quite a powerful 
tool to investigate the thermodynamics of several spin models and electron 
systems \cite{JWS}-\cite{tsuboi}. In most cases, after applying the BA
it proved useful 
to introduce {\it auxiliary functions} which 
are ratios of the components of the eigenvalue of the QTM 
and their non-linear integral equation (NLIE). 
The main merit to utilize the NLIE is that we do not have to use 
the string hypothesis \cite{string}, 
which sometimes raises fundamental questions about its validity as well as 
pragmatical questions about the accuracy of the unavoidable
truncation procedures of the typically 
infinitely many integral equations in the traditional 
thermodynamical Bethe ansatz (TBA) \cite{string}.

Once discovering a good (finite) set of auxiliary functions, we can get 
as good an accuracy as we need, because no truncation is necessary.
Several remarkable studies of the thermodynamics \cite{kl,GAJ1,GAJ2} 
and the mathematical aspects \cite{GAJ3} have been achieved by this scheme. 

In this paper, we consider the one-dimensional 
Uimin-Sutherland (US) model \cite{US}. For $m$ fermionic and $n$ bosonic
components the system is referred to as the $(m,n)$ model. For this general 
case the quantum transfer matrix is constructed and its eigenvalue 
equations are derived.
The $(3,0)$-US model is studied in detail for which the {\it spins} 
attached to the sites take three states, which can be regarded 
as the three vectors of the vector representation of the Lie algebra $sl(3)$. 
We have succeeded in finding the NLIE for the $(3,0)$-US model 
with regular analyticity properties. To obtain these well-posed NLIE for 
the $(3,0)$-US model, it is crucial to consider the anti-symmetrical
fusion (ASF) model in addition to the defining one.
The ASF of the vector representation of $sl(3)$ is another three-dimensional 
one, and thus called {\it conjugate}. The spinon like picture in the 
$(3,0)$-US model 
is made clear with the help of the NLIE. As an application some physical 
quantities like specific heats are calculated with good accuracy by
numerical treatments of the NLIE.  

The remainder of this paper is organized as follows. 
We give an explanation of the US model and the QTM method in Sec.2. 
This section is also intended to fix notations.
In Sec.3, we apply the BA to the QTM of the general US model and we also 
introduce the NLIE for the (3,0) model.
In several limiting cases, for example the $SU(2)$-limit, 
exact analytic calculations are performed. In Sec.4, we solve 
the NLIE numerically and obtain the entropy, specific heat, magnetization, 
and magnetic susceptibility. Some low-temperature
properties exposed by these data are discussed. 
In Sec.5 we give a summary of our work.

\section{The Uimin-Sutherland model and its QTM}
Let us begin with the review and definition of the general US models.  
The general one-dimensional $q$-state US 
model is defined as follows. Consider a one-dimensional lattice with $L$ 
sites and periodic boundary conditions imposed. A $q$-state spin 
variable $\alpha_{i}$ is assigned to each site $i$. 
We can generally consider the situation where each spin $\alpha$ has 
its own grading, i.e. statistics number 
$\epsilon_{\alpha}=(-1)^{\xi_{\alpha}}=\pm 1$. A spin $\alpha$ with 
$\epsilon_{\alpha}=+1$ ($\epsilon_{\alpha}=-1$) is called 
bosonic (fermionic). The Hamiltonian of the US model  
can be introduced as 
\BE
{\cal H}_{0}=\sum_{i=1}^{L}\pi_{i,i+1} \label{eq:UShamiltonian}
\EE
with the permutation operator $\pi_{i,i+1}$
\[
\pi_{i.i+1}\ket{\alpha_{1}\cdots\alpha_{i}\alpha_{i+1}\cdots\alpha_{L}}
=(-1)^{\xi_{\alpha_{i}\alpha_{i+1}}}
\ket{\alpha_{1}\cdots\alpha_{i+1}\alpha_{i}\cdots\alpha_{L}},
\]
where $\xi_{\alpha_{i}\alpha_{i+1}}$ is $1$ if both $\alpha_{i}$ and 
$\alpha_{i+1}$ are fermionic, and $0$ otherwise.

Model (\ref{eq:UShamiltonian})
is shown to be exactly solvable on the basis 
of the Yang-Baxter equation. Many well-known exactly solvable models are 
of type (\ref{eq:UShamiltonian}), e.g. the
spin-$1/2$ Heisenberg chain corresponds to $q=2$ and 
$\epsilon_{1}=\epsilon_{2}=+1$, the free fermion model to 
$q=2$ and $\epsilon_{1}=-\epsilon_{2}=+1$, the supersymmetric \tJ 
model to $q=3$ and $\epsilon_{1}=-\epsilon_{2}=\epsilon_{3}=+1$. 
If $m$ of $q$ $\epsilon$'s are $+1$ and $n (=q-m)$ are $-1$, for example, 
$\epsilon_{1}=\cdots=\epsilon_{m}=+1$, $\epsilon_{m+1}=\cdots\epsilon_{q}=-1$, 
we call the model $(m,n)$-US model.

Now we want to consider the $(3,0)$-US model 
($q=3,\,\epsilon_{1}=\epsilon_{2}=\epsilon_{3}=+1$)
whose Hamiltonian is equivalent to that of the $SU(2)$ spin-1 chain defined by 
\[
{\cal H_{0}}=\sum_{i=1}^{L}
\left[
{\vec S}_{i}\cdot{\vec S}_{i+1}+({\vec S}_{i}\cdot{\vec S}_{i+1})^{2}
\right].
\]

For the study of the thermodynamics we introduce the QTM as follows.
First, the classical counterpart to (\ref{eq:UShamiltonian}) the
Perk-Schultz (PS) model \cite{PS} is defined on a two-dimensional 
square lattice of $L\times N$ sites, where we impose periodic boundary 
conditions throughout this paper. 
We assume that variables taking on values 
$1,2,\cdots,q$ are assigned to the bonds of the lattice. The Boltzmann 
weight associated with a local vertex configuration $\alpha$, $\beta$, $\mu$ 
and $\nu$ is denoted by $R^{\mu\nu}_{\alpha\beta}(v)$, 
where $v$ is the spectral parameter (Fig.1).
%

%
%
Using the Yang-Baxter equation, 
a lattice model defined by
\begin{equation}
R^{\mu\nu}_{\alpha\beta}(v)=\delta_{\alpha\nu}\delta_{\mu\beta}+
v\cdot (-1)^{\xi_{\alpha}\xi_{\mu}}\cdot\delta_{\alpha\beta}\delta_{\mu\nu}
\label{eq:PS-model}
\end{equation}
is proved to be exactly solvable.
Defining the row-to-row  
transfer matrix  
\[
\fT^{\beta}_{\alpha}(v) = \sum_{\{ \mu\} }
                           \prod_{i=1}^{L}
                           R_{\alpha_{i}\beta_{i}}^{\mu_{i}\mu_{i+1}}(v), 
\]
the partition function is given by
${\bf {\sf Z}}_{L,N}= Tr\fT^{N}(v)$
where the trace is taken in the $q^{L}$-dimensional space. In this paper
we are not primarily interested in the PS-model itself, but rather in its
Hamiltonian limit obtained from $\fT$. 
Making use of Baxter's formula \cite{baxterbook} at $v=0$
\BE
\fH_{0}=\left. {d\over dv}\ln\fT(v)\right|_{v=0}
=\sum_{i=1}^{N}\fh_{i,i+1}, \label{eq:baxter}
\EE
we get the Hamiltonian of the $q$-state US model. 

The main idea of the quantum transfer matrix (QTM) method at finite 
temperature is as simple as follows (for details the reader is referred to the 
papers \cite{kl}). 
First, let us define a new set of
vertex weights ${\bar R}(v)$ by rotating $R(v)$ by 90 degrees as  
\[
{\bar R}_{\alpha\beta}^{\mu\nu}(v)=R_{\nu\mu}^{\alpha\beta}(v)
\]
and consider the transfer-matrix ${\bar\fT}$ which is the product of 
${\bar R}(v)$. 
Upon introducing 
a large integer $N$ (Trotter number), we get the following relation 
with arbitrary reciprocal temperature $\beta$
\[
\fT(-\beta/N){\bar\fT}(-\beta/N)=
e^{-{2\beta\over N}\fH+O((\beta/N)^{2})}.
\]
Finally, the partition function of the one-dimensional Hamiltonian $\fH$ 
at finite temperature $T=1/\beta$ can be calculated by means of the
following ``Trotter-Suzuki'' formula
\BE
{\sf Z}=\Tr e^{-\beta\fH}
=\lim_{N\rightarrow\infty}\LR\Tr\fT(u){\bar\fT}(u)\RR^{N/2}, 
\quad u=-\beta/N.\label{eq:trotter}\EE
In other words, the finite-temperature partition function for the 
one-dimensional Hamiltonian is calculated as that of 
a {\em staggered}
two-dimensional vertex model. For technical reasons, it is 
convenient
to define another vertex weight ${\tilde R}(v)$ as a rotation
of $R(v)$ by -90 degrees
\[
{\tilde R}_{\alpha\beta}^{\mu\nu}(v)=R_{\mu\nu}^{\beta\alpha}(-v).
\]
With these preparations, the quantum transfer matrix $\fT_{\rm QTM}$ 
corresponding to the contribution of columns to the partition function
is given by
\BE
\LR\fT^{\rm QTM}\RR^{\beta}_{\alpha}(v)=\sum_{\mu}\prod_{j=1}^{N/2}
R_{\alpha_{2j-1}\beta_{2j-1}}^{\mu_{2j-1}\mu_{2j}}(iv+u)
{\tilde R}_{\alpha_{2j}\beta_{2j}}^{\mu_{2j}\mu_{2j+1}}(iv-u).
\EE
Here we have introduced a spectral parameter $v$ such that $\fT_{\rm QTM}(v)$ 
is a commuting family of matrices generated by $v$. This will allow us to 
diagonalize $\fT_{\rm QTM}$. The final results, of course, are physically
interesting only for $v=0$ as the partition function of the 
one-dimensional US model at 
temperature $1/\beta$ is given by
\[
\fZ=\lim_{N\rightarrow\infty}\Tr\fT_{\rm QTM}^{L}(0).
\]
The free energy per unit length is
\[
f=-\lim_{L\rightarrow\infty}{1\over L\beta}\ln{\sf Z}
=-{1\over\beta}\ln\Lambda_{\rm max}(0),
\]
where $\Lambda_{\rm max}(v)$ is the largest eigenvalue of the 
QTM. If we succeed in obtaining the next-largest 
eigenvalue $\Lambda_{1}(v)$, the correlation length $\xi$ at the 
finite temperature $T=1/\beta$ is given by 
\[
\xi^{-1}=-\lim_{N\rightarrow\infty}\ln
\left|{\Lambda_{1}\over \Lambda_{\rm max}}\right| .
\]

\section{Nonlinear Integral Equation for the Free Energy}
Let us apply the BA for this QTM. 
While keeping solvability, we can add external field 
terms $\fH_{\rm ext}$ to $\fH_{0}$ like 
\[
\fH=\fH_{0}+\fH_{\rm ext}=
\fH_{0}-\sum_{i=1}^{L}\sum_{\alpha=1}^{q}
\mu_{\alpha}n_{i,\alpha}.
\]
The BA for the eigenvalues of 
the QTM of the general $q$-state PS model is conjectured to take the form
of
\BEA
\Lambda(v)&=&\sum_{j=1}^{q}\lambda_{j}(v), \NN \\
\lambda_{1}(v)&=&
\prod_{k_{1}=1}^{M_{1}}{v-v_{k_{1}}^{1}+i\epsilon_{1}\over v-v_{k_{1}}^{1}}
(v-iu-i\epsilon_{1})^{N\over 2}(v+iu)^{N\over 2}e^{\beta\mu_{1}}, \NN \\
\lambda_{j}(v)&=&
\prod_{k_{j-1}=1}^{M_{j-1}}
{v-v_{k_{j-1}}^{j-1}-i\epsilon_{j}\over v-v_{k_{j-1}}^{j-1}}
\prod_{k_{j}=1}^{M_{j}}{v-v_{k_{j}}^{j}+i\epsilon_{j}\over v-v_{k_{j}}^{j}}
(v-iu)^{N\over 2}(v+iu)^{N\over 2}e^{\beta\mu_{j}},\NN \\
&&\hskip10cm (j=2,\cdots,q-1),\NN \\
\lambda_{q}(v)&=&
\prod_{k_{q-1}=1}^{M_{q-1}}
{v-v_{k_{q-1}}^{q-1}-i\epsilon_{q}\over v-v_{k_{q-1}}^{q-1}}
(v+iu+i\epsilon_{q})^{N\over 2}(v
-iu)^{N\over 2}e^{\beta\mu_{q}}.
\EEA
We concentrate on the $(3,0)$-US model. 
Defining $\fq_{i}(v)=\prod(v-v_{k_{i}}^{i})$ and 
$\phi_{\pm}(v)=(v\pm iu)^{N/2}$, 
the last formulas are reduced to
\begin{eqnarray}
&&\lambda_{1}(v)={\fq_{1}(v+i)\over \fq_{1}(v)}\phi_{+}(v)\phi_{-}(v-i)
e^{\beta\mu_{1}},\NN\\
&&\lambda_{2}(v)={\fq_{1}(v-i)\over \fq_{1}(v)}{\fq_{2}(v+i)\over \fq_{2}(v)}
\phi_{+}(v)\phi_{-}(v)e^{\beta\mu_{2}},\NN\\
&&\lambda_{3}(v)={\fq_{2}(v-i)\over \fq_{2}(v)}\phi_{+}(v+i)\phi_{-}(v)
e^{\beta\mu_{3}},\NN\\
&&\Lambda(v)=\lambda_{1}(v)+\lambda_{2}(v)+\lambda_{3}(v).
\end{eqnarray}
We have checked numerically for small $N$'s 
that the above BA gives the correct eigenvalues.

Next, we want to define auxiliary functions which determine
the (largest) eigenvalue completely and satisfy a closed system of integral
equations.
In the case of the \tJ model, a complete set of
auxiliary functions is found \cite{GAJ1} to be 
\BE
b={\lambda_{1}\over\lambda_{2}+\lambda_{3}},\,
{\bar b}={\lambda_{3}\over\lambda_{1}+\lambda_{2}},
c={\lambda_{1}\lambda_{3}\over
   \lambda_{2}(\lambda_{1}+\lambda_{2}+\lambda_{3})}.
\EE 
Unfortunately, for the $(3,0)$-US model the above set does not admit a closed
set of functional equations to determine the eigenvalue. 
In order to complete this set we found
in the fusion procedure \cite{fusion} a useful working basis
\cite{GAJ2,tsuboi}. 
The fusion model 
is defined as follows. The PS model given by (\ref{eq:PS-model}) is 
identified with the vector representation of the $sl(m|n)$ model. On the basis
of tensor products and proper projections, we get the other representations 
with higher levels. The solvability of the fusion model is
again guaranteed by 
the Yang-Baxter relation. 
Here we consider only the ASF model. 
For the case of $sl(n)$, we call the $(n-1)$-th 
ASF model the {\it conjugate model} because the ASF 
of the fundamental model and the $(n-1)$-th ASF is 
zero-dimensional. This conjugacy is somewhat like that between 
quark and anti-quark representations of $SU(3)$. 
For the $sl(2)$ case, the ASF of two vector representations 
({\it i.e} spin-$1/2$) gives the zero-dimensional representation 
({\it i.e} spin-$0$). Therefore, we call the $sl(2)$ PS model self-conjugate.  
Under the conjugate transformation, we see 
\[
\hitotu{1}\leftrightarrow\hitotu{2}\quad .
\]
For the $sl(3)$ case, 
\[
\hitotu{1}\leftrightarrow\hutatu{2}{3}\, ,\,
\hitotu{2}\leftrightarrow\hutatu{1}{3}\, ,\,
\hitotu{3}\leftrightarrow\hutatu{1}{2}\, .
\]
The BA equation for the fusion model is obtained by the 
simple replacement
\BEA
\lambda_{1}(v)\rightarrow\lambda_{2,3}(v),\, 
\lambda_{2}(v)&\rightarrow&\lambda_{1,3}(v),\, 
\lambda_{3}(v)\rightarrow\lambda_{1,2}(v), \\
\Lambda(v)\rightarrow{\tilde \Lambda}(v)&=&
\lambda_{2,3}(v)+\lambda_{1,3}(v)+\lambda_{1,2}(v),
\EEA
where $\lambda_{l,m}(v)$ is defined by
\[
\lambda_{l,m}(v)=\lambda_{l}(v)\lambda_{m}(v+i).
\]

We are ready to define the auxiliary functions. 
In the case of the $sl(2)$ PS model, 
the ASF model is nothing but the fundamental model itself. As is known, 
a complete set of auxiliary functions consists of 
$p=\lambda_{1}/\lambda_{2}$ and ${\bar p}=p^{-1}=\lambda_{2}/\lambda_{1}$. 
In fact, $p$ and ${\bar p}$ are conjugates. For the $sl(3)$ case, 
the three functions $b$, ${\bar b}$ and $c$ are not complete 
in the $(3,0)$-US model, however
the 6 functions $b$, ${\bar b}$, $c$ and their conjugates 
constitute a complete set. For this reason
we have introduced the conjugate transformation. Let us proceed this program. 
In dependence on the real variable $x$, we define the functions 
\begin{eqnarray}
s_{1}(x)=\suka{\lambda_{1}\over \lambda_{2}+\lambda_{3}}\bou{v=x+i/2}&,&
s_{2}(x)=\suka{\lambda_{12}\lambda_{23}
      \over \lambda_{13}(\lambda_{12}+\lambda_{23}+\lambda_{13})}
      \bou{v=x-i/2},\NN \\
s_{3}(x)=\suka{\lambda_{3}\over \lambda_{1}+\lambda_{2}}\bou{v=x-i/2}&,&
s_{4}(x)=\suka{\lambda_{12}\over \lambda_{13}+\lambda_{23}}\bou{v=x},\\
s_{5}(x)=\suka{\lambda_{1}\lambda_{3}
      \over \lambda_{2}(\lambda_{1}+\lambda_{2}+\lambda_{3})}\bou{v=x}&,&
s_{6}(x)=\suka{\lambda_{23}\over \lambda_{12}+\lambda_{13}}\bou{v=x-i},\NN\\
\Lambda(x)=\suka\lambda_{1}+\lambda_{2}+\lambda_{3}\bou{v=x}&,&
{\bar \Lambda}(x)=\suka\lambda_{12}+\lambda_{23}+\lambda_{13}\bou{v=x-i/2}.\NN
\end{eqnarray}
After a lengthy calculation using the Fourier transform
the following non-linear integral equation are proved explicitly
\begin{equation}
\left(
\begin{array}{c}
\ln s_{1}(x)\\ \ln s_{2}(x)\\ \ln s_{3}(x)\\ 
\ln s_{4}(x)\\ \ln s_{5}(x)\\ \ln s_{6}(x) 
\end{array}
\right)
=
-
\left(
\begin{array}{c}
\beta\epsilon_{1}(x)\\ \beta\epsilon_{2}(x)\\ \beta\epsilon_{3}(x)\\ 
\beta\epsilon_{4}(x)\\ \beta\epsilon_{5}(x)\\ \beta\epsilon_{6}(x)
\end{array}
\right)
+
\left(
\begin{array}{cccccc}
 K_{0}&-K_{1}&-K_{1}&-K_{3}&-K_{3}&-K_{4}\\
-K_{2}& K_{0}&-K_{1}&-K_{3}&-K_{6}&-K_{3}\\
-K_{2}&-K_{2}& K_{0}&-K_{5}&-K_{3}&-K_{3}\\
-K_{3}&-K_{3}&-K_{4}& K_{0}&-K_{1}&-K_{1}\\
-K_{3}&-K_{6}&-K_{3}&-K_{2}& K_{0}&-K_{1}\\
-K_{5}&-K_{3}&-K_{3}&-K_{2}&-K_{2}& K_{0}
\end{array}
\right)
*
\left(
\begin{array}{c}
\ln S_{1}(x)\\ \ln S_{2}(x)\\ \ln S_{3}(x)\\ 
\ln S_{4}(x)\\ \ln S_{5}(x)\\ \ln S_{6}(x)
\end{array}
\right).
\label{eq:NLIE}
\end{equation}
Several remarks are in order. First, $S_{i}=1+s_{i}\quad(i=1,\cdots,6)$. 
Second, the driving terms/bare energies of the {\it spinons} 
are defined as
\begin{eqnarray}
\epsilon_{1}(x)&=&V_{1}(x)+(-2\mu_{1}+\mu_{2}+\mu_{3})/3, \\
\epsilon_{2}(x)&=&V_{1}(x)+(\mu_{1}-2\mu_{2}+\mu_{3})/3,\\
\epsilon_{3}(x)&=&V_{1}(x)+(\mu_{1}+\mu_{2}-2\mu_{3})/3, \\
\epsilon_{4}(x)&=&V_{2}(x)+(-\mu_{1}-\mu_{2}+2\mu_{3})/3, \\
\epsilon_{5}(x)&=&V_{2}(x)+(-\mu_{1}+2\mu_{2}-\mu_{3})/3,\\
\epsilon_{6}(x)&=&V_{2}(x)+(2\mu_{1}-\mu_{2}-\mu_{3})/3
\end{eqnarray}
with
\BE
V_{1}(x)={2\pi\over\sqrt{3}}{1\over 2\cosh(2\pi x/3)-1},\quad 
V_{2}(x)={2\pi\over\sqrt{3}}{1\over 2\cosh(2\pi x/3)+1}.
\EE
Lastly, the kernels $K_{l}(x)$($l=0,\cdots,6$) are 
\[
K_{l}(x)=\IFT K_{l}(k)
\]
with
\begin{eqnarray}
&&K_{0}(k)={e^{-|k|}\over e^{k}+1+e^{-k}},\,
K_{1}(k)={1+e^{-3k/2-|k|/2}\over e^{k}+1+e^{-k}},\,
K_{2}(k)={1+e^{ 3k/2-|k|/2}\over e^{k}+1+e^{-k}}, \NN \\
&&K_{3}(k)={e^{|k|/2}\over e^{k}+1+e^{-k}},\,
K_{4}(k)={e^{-3k/2-|k|}\over e^{k}+1+e^{-k}},
K_{5}(k)={e^{ 3k/2-|k|}\over e^{k}+1+e^{-k}}, \\
&&K_{6}(k)={e^{-|k|/2}+2e^{|k|/2}+e^{-3|k|/2}\over e^{k}+1+e^{-k}}, \NN
\end{eqnarray}
and the convolution is defined by
\[
f*g(x)=\int_{-\infty}^{\infty}{dy\over 2\pi}f(x-y)g(y).
\]
We introduce the rescaled eigenvalue 
$\Lambda'(v) = \Lambda(v)/[\phi_{+}(v+i)\phi_{-}(v-i)]$ which is useful,
because of the simple asymptotic behaviour 
$\displaystyle{\lim_{v\rightarrow\infty}}\Lambda'(v)=$ const. At $v=0$, the 
relation of eigenvalue and rescaled eigenvalue is 
simply
\[
\ln\Lambda(0)=\ln\Lambda'(0)-\beta,
\]
i.e. it only amounts to a shift in the ground state energy.
In terms of the auxiliary functions the rescaled eigenvalue reads
\BEA
\ln \Lambda'(x)=-\beta e(x)&+&
V_{1}*\ln S_{1}(x)+V_{1}*\ln S_{2}(x)+V_{1}*\ln S_{3}(x)\NN \\
&+& V_{2}*\ln S_{4}(x)+V_{2}*\ln S_{5}(x)+V_{2}*\ln S_{6}(x)\NN
\EEA
with 
\[
e(0)=-\int dk{1+e^{-|k|}\over e^{k}+1+e^{-k}}=
-\left({\pi\over 3\sqrt{3}}+\ln 3\right).
\]
From considering the low temperature limit by following \cite{kl}, 
we conclude that there are two kinds of elementary excitations.  
One of them corresponds to the vector representation $(s_{1},s_{3},s_{5})$, 
and the other to 
its conjugate $(s_{2},s_{4},s_{6})$. 
Their bare energies are represented by 
$\epsilon_{1},\cdots\epsilon_{6}$. 

\newpage
\noindent
{\large\bf Limiting cases}

Next, we take a suitable limit of the chemical 
potentials $\mu_{1}$, $\mu_{2}$ and $\mu_{3}$, for which
the spin-1/2 Heisenberg 
model is obtained. It is quite a natural result on the level of the
Hamiltonian formalism, however less trivial on the level of the final
NLIE, which supports the validity of our construction and the correctness
of our calculations.
Let us concentrate on the case of 
$\mu_{1}=\mu'+h_{1}$,  $\mu_{3}=\mu'+h_{3}$ and $\mu_{2}=h_{2}$, where 
$|h_{1}|,|h_{2}|,|h_{3}|\ll |\mu'|$.  
First consider the limit $\mu'=\infty$. In this case, observing the 
constants in the driving terms of each auxiliary function, we conclude
\begin{eqnarray}
s_{1} = O(1),&s_{2} = O(e^{-2\beta\mu'}),&
s_{3} = O(1),\\
s_{4} = O(e^{-\beta\mu'}),
& s_{5} = O(e^{\beta\mu'}),&s_{6} = O(e^{-\beta\mu'}).
\\ \NN
\end{eqnarray} 
Therefore, we can regard
\begin{eqnarray}
s_{2} = s_{4} &=& s_{6} = 0,\NN \\
s_{5} &\sim& S_{5}
\end{eqnarray}
that means the conjugate modes are suppressed.
With the above ansatz, we can solve the equation for the $S_{2}$-function as
\[
\ln S_{5}(x)=K_{3}*(K_{0}-1)^{-1}*(\ln S_{1}(x)+\ln S_{3}(x))
+\beta(K_{0}-1)^{-1}*V_{2}(x)
\]
Substituting the last equation into (\ref{eq:NLIE})
for $s_{1}$ and $s_{3}$ and 
performing the inverse Fourier transform, we get 
\begin{eqnarray}
\ln p(x+i/2)&=&
-2\pi\beta\Phi(x+i/2)+\beta h/2 \NN\\
&&\qquad +\int_{-\infty}^{\infty}{dy\over 2\pi}k(x-y)\ln P(y+i/2)
-\int_{-\infty}^{\infty}{dy\over 2\pi}k(x-y+i)\ln{\bar P}(y-i/2), \NN\\
\ln{\bar p}(x-i/2)&=&
2\pi\beta\Phi(x-i/2)-\beta h/2 \\
&&\qquad-\int_{-\infty}^{\infty}{dy\over 2\pi}k(x-y-i)\ln P(y+i/2)
+\int_{-\infty}^{\infty}{dy\over 2\pi}k(x-y)\ln{\bar P}(y-i/2), \NN\\
\ln\Lambda(x)&=&2\beta\ln 2+\beta\mu+{i\over 2}\int_{-\infty}^{\infty} dx
{\ln P(x+i/2)\over \sinh\pi(x+i/2)}-{i\over 2}\int_{-\infty}^{\infty} dx
{\ln{\bar P}(x-i/2)\over \sinh\pi(x-i/2)}\NN
\end{eqnarray} 
with
\begin{eqnarray}
&&p(x+i/2)=s_{1}(x),\,{\bar p}(x-i/2)=s_{3}(x),\,
P(v)=1+p(v),\,{\bar P}(v)=1+{\bar p}(v),\NN\\
&&\Phi(v)=-{i\over 2}{1\over \sinh\pi v},\,
k(x)=\int{dx\over 2\pi}{e^{ikx}\over 1+e^{|k|}},\,
h=h_{1}-h_{3},\, \mu={h_{1}+h_{2}+h_{3}\over 3},
\end{eqnarray}
where we have dropped the contribution of $\mu'$ to
the potential $\mu$.
This NLIE is nothing but that for the spin 1/2-Heisenberg model \cite{kl}. 
Thus, we see that the spin 1/2-Heisenberg model is obtained
as a limit of the $(3,0)$-US model directly in the two sets of NLIE's.   

Now, let us consider the opposite limit $\mu'=-\infty$. In this case, we 
find the following simplifications
\begin{eqnarray}
s_{1} = s_{3} &=& s_{5} = 0,\NN \\
s_{2} &\sim& S_{2}.
\end{eqnarray}
Therefore, we obtain
\begin{equation}
\ln S_{2}(x)=K_{3}*(K_{0}-1)^{-1}*(\ln S_{4}(x)+\ln S_{6}(x))
+\beta(K_{0}-1)^{-1}*V_{1}(x). \label{eq:limit2}
\end{equation}
However, substituting (\ref{eq:limit2}) into (\ref{eq:NLIE}), 
we see that $s_{4}$, $s_{6}$ and $\Lambda$ are constants. 
This fact is reasonable, because in the present case only the 2nd 
state can survive with finite energy. Hence, all physical degrees of freedom
are frozen out at finite temperature.

\section{Numerical Analysis of the NLIE}
As an application of the obtained NLIE (\ref{eq:NLIE}), 
we show numerical results for
some physical quantities. 
Let us consider the entropy and the specific heat:
\begin{equation}
S=-\left({\partial f\over\partial T}\right),\quad 
C=T\left({\partial S\over\partial T}\right).
\end{equation}
To avoid numerical differentiations, we simultaneously 
have solved the NLIE for the derivatives using 
relations like \cite{GAJ1}
\[
{\partial\over\partial\beta}\ln(1+s_{i})=
{s_{i}\over 1+s_{i}}{\partial\over\partial\beta}\ln s_{i}.
\] 
We have calculated the entropy and specific heat of the 
$(3,0)$-US model with $\mu_{1}=\mu_{2}=\mu_{3}=0$ numerically as 
shown in Fig.2. 

We can compare the result with that from the $SL(3)_{1}$ 
conformal field theory (CFT). 
From CFT and its finite size analysis, 
the free energy is predicted to be
\begin{equation}
f=f_{0}-{\pi c\over 6v}T^{2}+\cdots. \label{eq:CFT}
\end{equation}
Putting $c=2$ and $v=2\pi/3$ \cite{devega}, the low-temperature asymptotics
is
\[
C = T.
\]
With a glance at Fig.2, we see that 
the low temperature behaviour of the numerically determined specific heat has 
slope $1$ as predicted by CFT. 
In a similar manner, we can calculate the specific heat in the 
presence of the chemical potential. First, we show results for the case
$\mu_{1}=\mu_{3}=\mu,\, \mu_{2}=0$ in Figs.3a and 3b. 
As we have analyzed before, 
the specific heat in this case approaches that of the 
$SU(2)$ spin-1/2 Heisenberg chain with zero magnetic field as 
$\mu\rightarrow\infty$. For the Heisenberg chain, we should put 
$c=1$ and $v=\pi$ and get the asymptotics $C=T/3$. Indeed, the 
curves in Fig.3a show the expected tendency.
Results for the specific heat for $\mu_{1}=\mu_{3}=\mu<0,\, \mu_{2}=0$ 
are shown in Fig.3b. 
Here we see a suppression of the specific heat data 
at fixed temperature for $\mu\to -\infty$. This is the manifestation of
the freezing of the system.
Second, the case of $\mu_{1}=-\mu_{3}=h,\, \mu_{2}=0$ is 
considered in Fig.4. 
In this case, the rich $sl(3)$ structure is most clearly exposed.

At $h=4$ in Fig.4, we observe two structures (one shoulder and
one maximum) of the specific heat owing to the fundamental mode and
its conjugate.

Let us define the magnetization $M$ and the magnetic
susceptibility $\chi$ by
\[
M=-{\partial f\over\partial h},\quad 
\chi=\left.{\partial M\over\partial h}\right|_{ h=0}
=-\left.{\partial^{2} f\over\partial h^{2}}\right|_{ h=0},
\]
respectively. The results are shown in Figs. 5a, b and 6. 
From Figs.5 we read off the existence of
two critical field $h_{c1}\cong 0.94$ and $h_{c2}=4$. 
For $h\geq h_{c2}$ only the $\alpha=1$ state can survive at $T=0$, thus
the groundstate magnetization is maximal, i.e. $M=1$. 
This is in agreement with analytic
considerations. The lower field $h_{c1}$ is the 
value, above which only the $\alpha=1$ and $\alpha=2$ states can exist. 
Note that for these critical fields the magnetization $M(T)$ shows a 
square root behaviour at low $T$.

The magnetic susceptibility is given in Fig.6. 
The curve is qualitatively similar to that of $SU(2)$ Heisenberg model
however with $\chi(0)=3/\pi^2$. The susceptibilities at the lowest 
temperatures shown in Fig. 6 are still about 10\% above the groundstate
value. The origin of this singular behaviour of the susceptibility at $T=0$
are $1/\log T$ corrections similar to \cite{log,Kl98} and will be 
discussed elsewhere. 
%
\section{Summary}
Before closing this paper, we would like to mention 
another application and generalizations of the 
NLIE. First, we can use our NLIE in order to calculate the characters of 
the $SL(3)_{1}$ Kac-Moody algebra. Let us remember that our NLIE 
contains a natural notion of spinons and chemical potentials
\cite{Haldane}-\cite{SUncharacter}. 
This fact enables us to present the characters in terms of the spinons 
and chemical potentials \cite{SUncharacterJun}. 
Details will be published elsewhere. Second, up to our knowledge
the complete strings in the anisotropic $sl(n)$ $(n>2)$ models
with trigonometric and elliptic $R$-matrices
have not been constructed yet. However, one of the present authors has shown 
that the QTM method is applicable not only to the isotopic $XXX$ 
Heisenberg chain but also to the anisotropic $XXZ$ and $XYZ$ versions
\cite{kl}. Therefore, we expect 
that a combination of the anti-symmetric fusion procedure and the 
NLIE will be generalizable to the anisotropic $sl(3)$ models.\\

{\bf Acknowledgments}\\
The authors would like to thank J. Suzuki for valuable discussions and 
a critical reading of the manuscript. They also acknowledge 
G. J{\"u}ttner for discussions and technical instructions. 
The authors  acknowledge  financial   support  by the   {\it  Deutsche
Forschungsgemeinschaft} under grant  No.   Kl~645/3-1 and support by
the research program of the 
Sonderforschungsbereich 341, K\"oln-Aachen-J\"ulich.

\newpage
\pagestyle{empty}
\Large{\bf{Figure Captions}}\\
~\\
\vspace*{1cm}
\begin{tabular}{ll}
Fig.1& $R$-matrix for the PS model.\\
&\\
Fig.2&  Entropy and specific heat in the absence of the chemical potentials.\\
&\\
Fig.3a&  Specific heat for $\mu_{1}=\mu_{3}=\mu\geq 0$.\\
&\\
Fig.3b&  Specific heat for $\mu_{1}=\mu_{3}=\mu\leq 0$.\\
&\\
Fig.4&  Specific heat with $\mu_{1}=-\mu_{3}=h\geq 0$.\\
&\\
Fig.5a&  Magnetization for $\mu_{1}=-\mu_{3}\cong h_{c2},\, \mu_{2}=0$.\\
&\\
Fig.5b&  Magnetization for $\mu_{1}=-\mu_{3}\cong h_{c1},\, \mu_{2}=0$.\\
&\\
Fig.6&  Magnetic susceptibility.
\end{tabular}
\newpage
\begin{figure}
  \epsfxsize = 8 cm   
  \centerline{\epsfbox{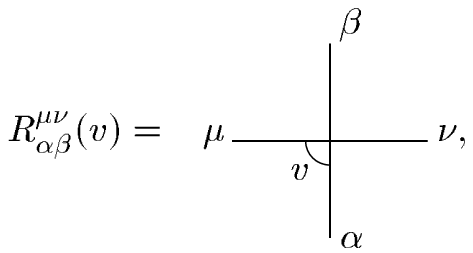}}
  \centerline{Fig.1}
\end{figure}
\begin{figure}
  \epsfxsize = 12 cm   
  \centerline{\epsfbox{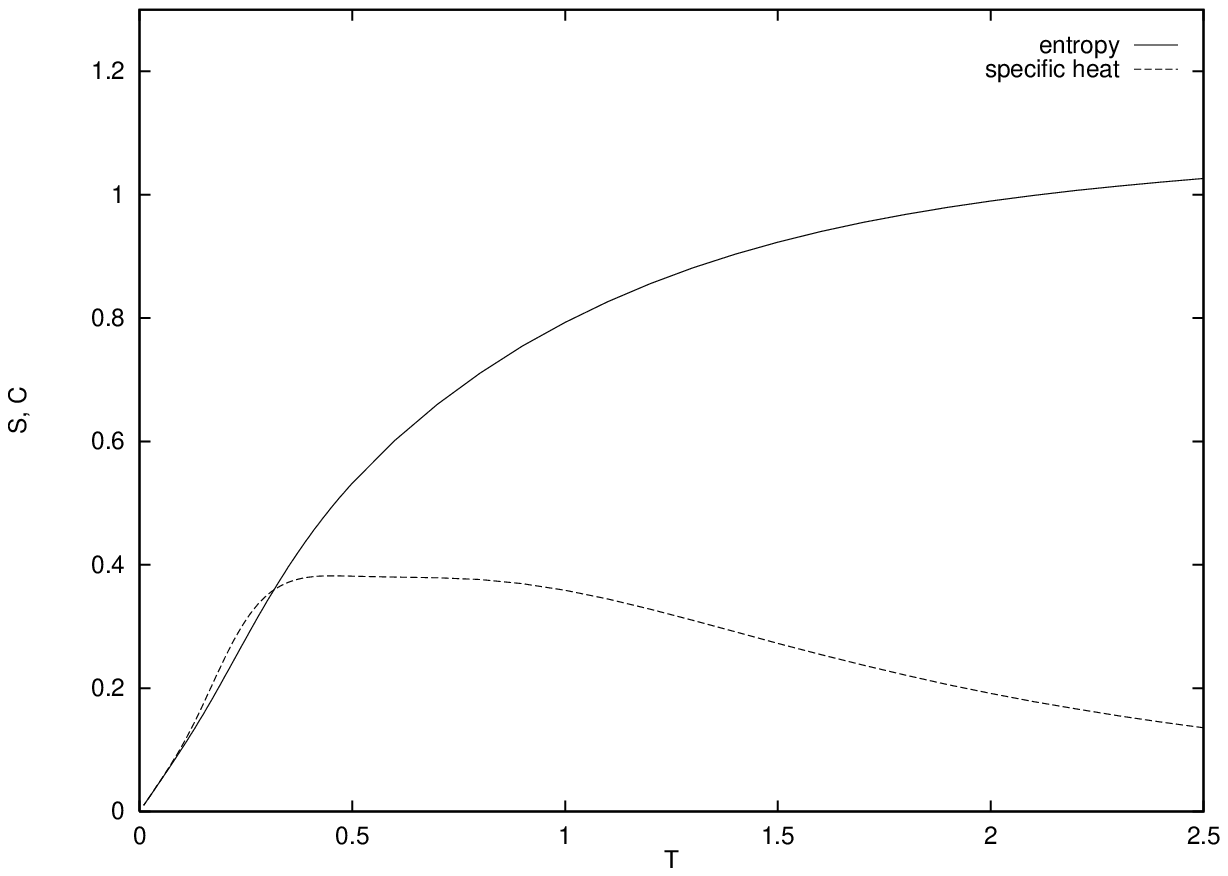}}
  \centerline{Fig.2}
\end{figure}
\newpage
\begin{figure}
  \epsfxsize = 12 cm   
  \centerline{\epsfbox{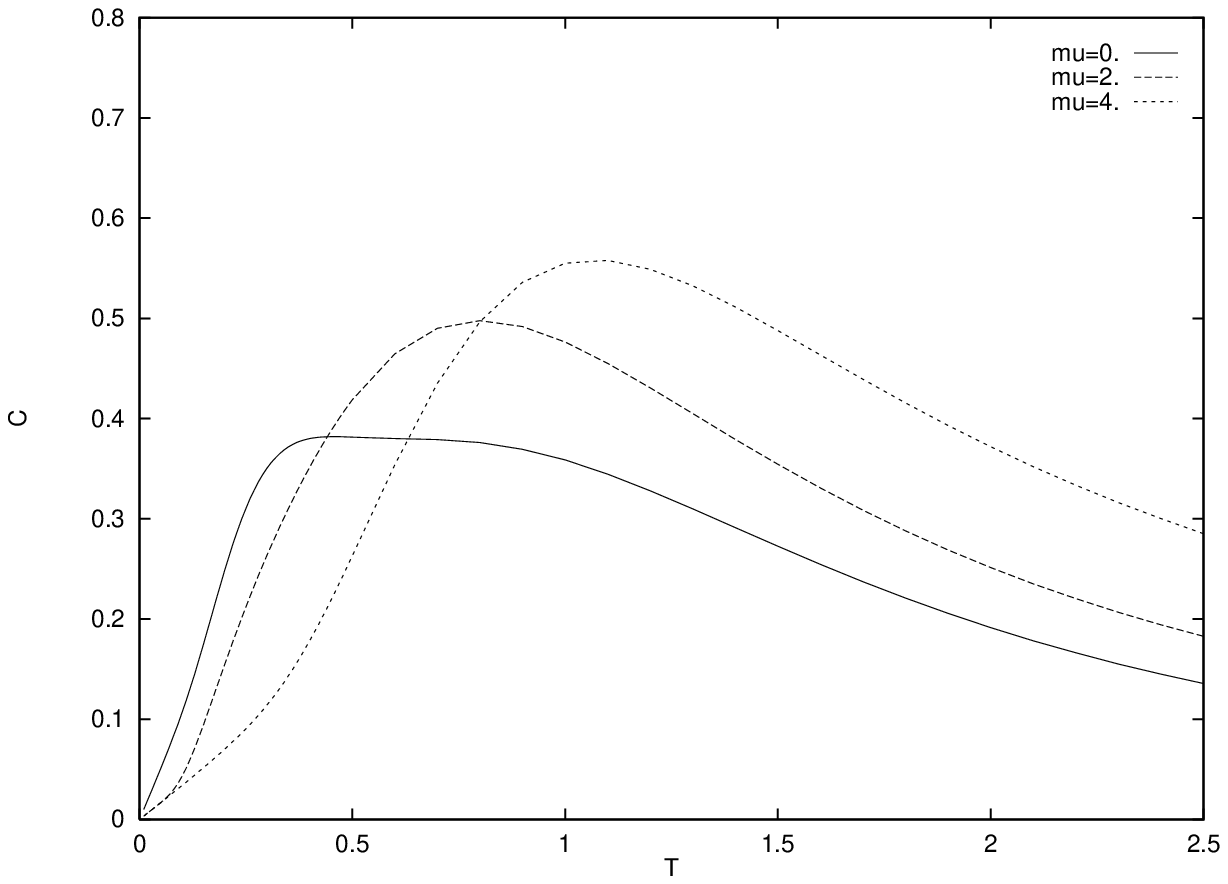}}
  \centerline{Fig.3a}
\end{figure}
\begin{figure}
  \epsfxsize = 12 cm   
  \centerline{\epsfbox{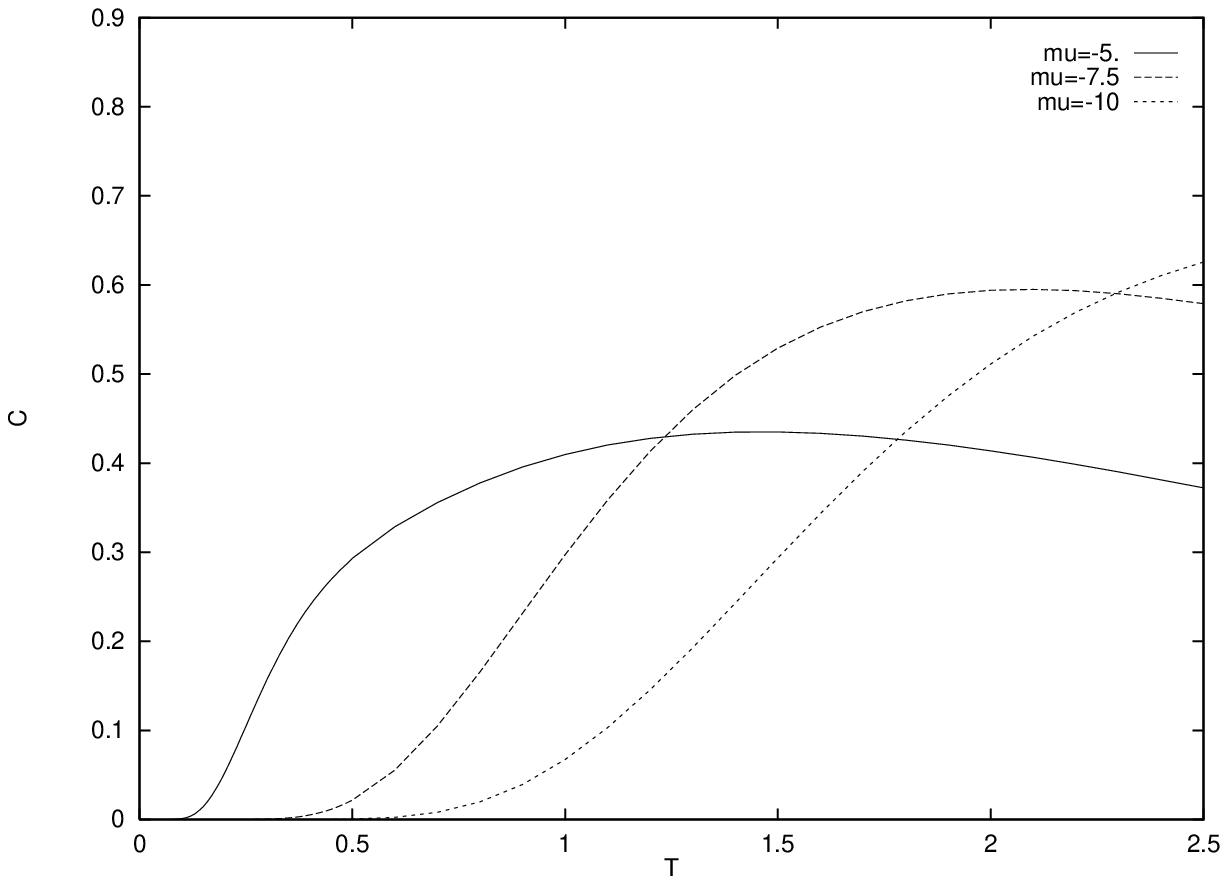}}
  \centerline{Fig.3b}
\end{figure}
\newpage
\begin{figure}
  \epsfxsize = 12 cm   
  \centerline{\epsfbox{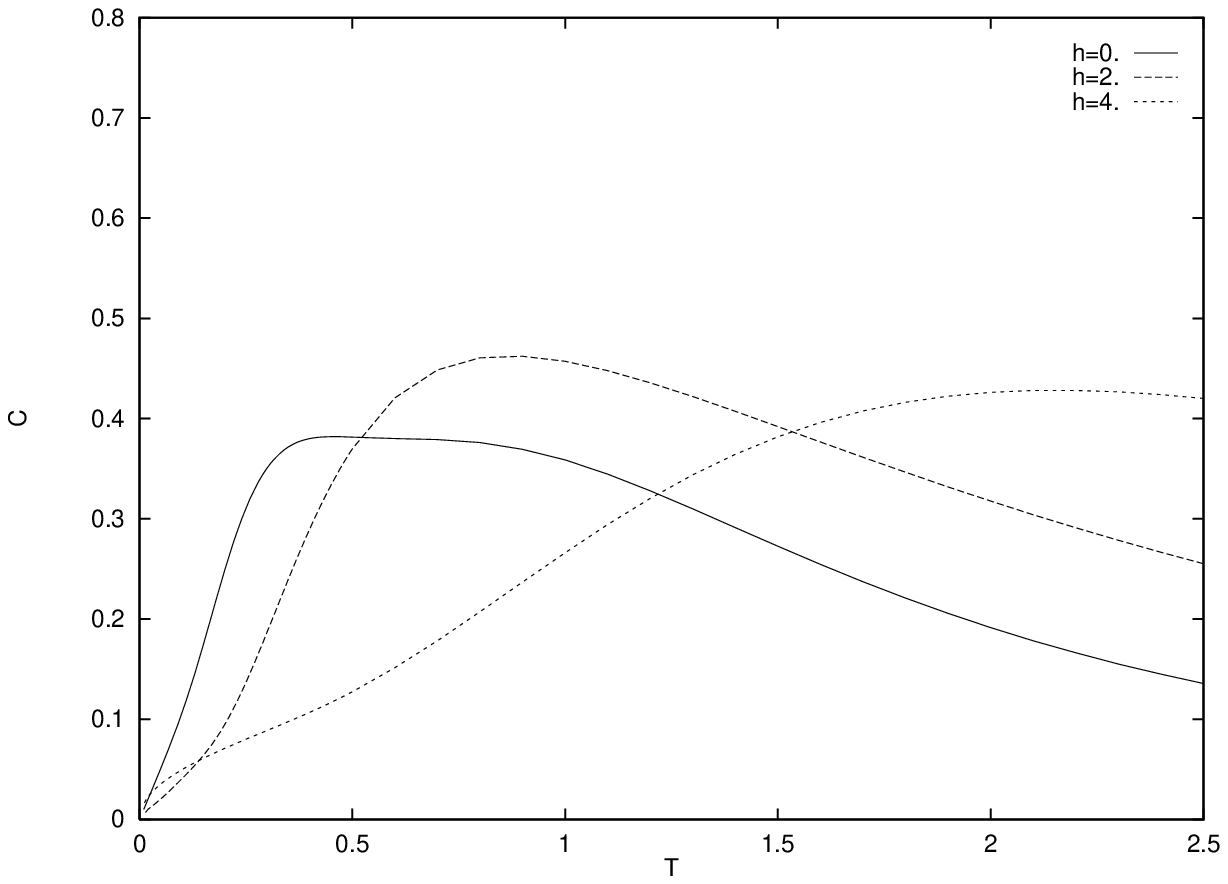}}
  \centerline{Fig.4}
\end{figure}
\newpage
\begin{figure}
  \epsfxsize = 12 cm   
  \centerline{\epsfbox{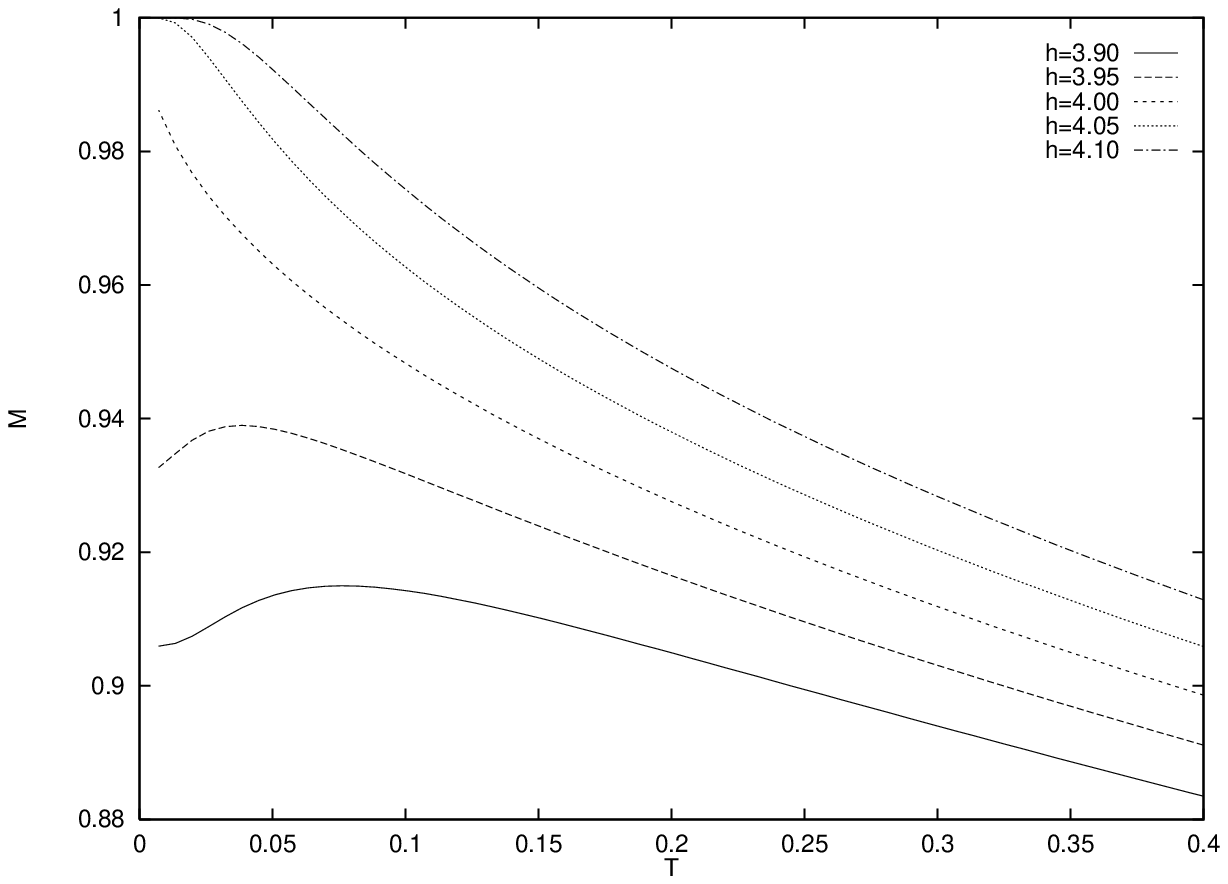}}
  \centerline{Fig.5a}
\end{figure}
\begin{figure}
  \epsfxsize = 12 cm   
  \centerline{\epsfbox{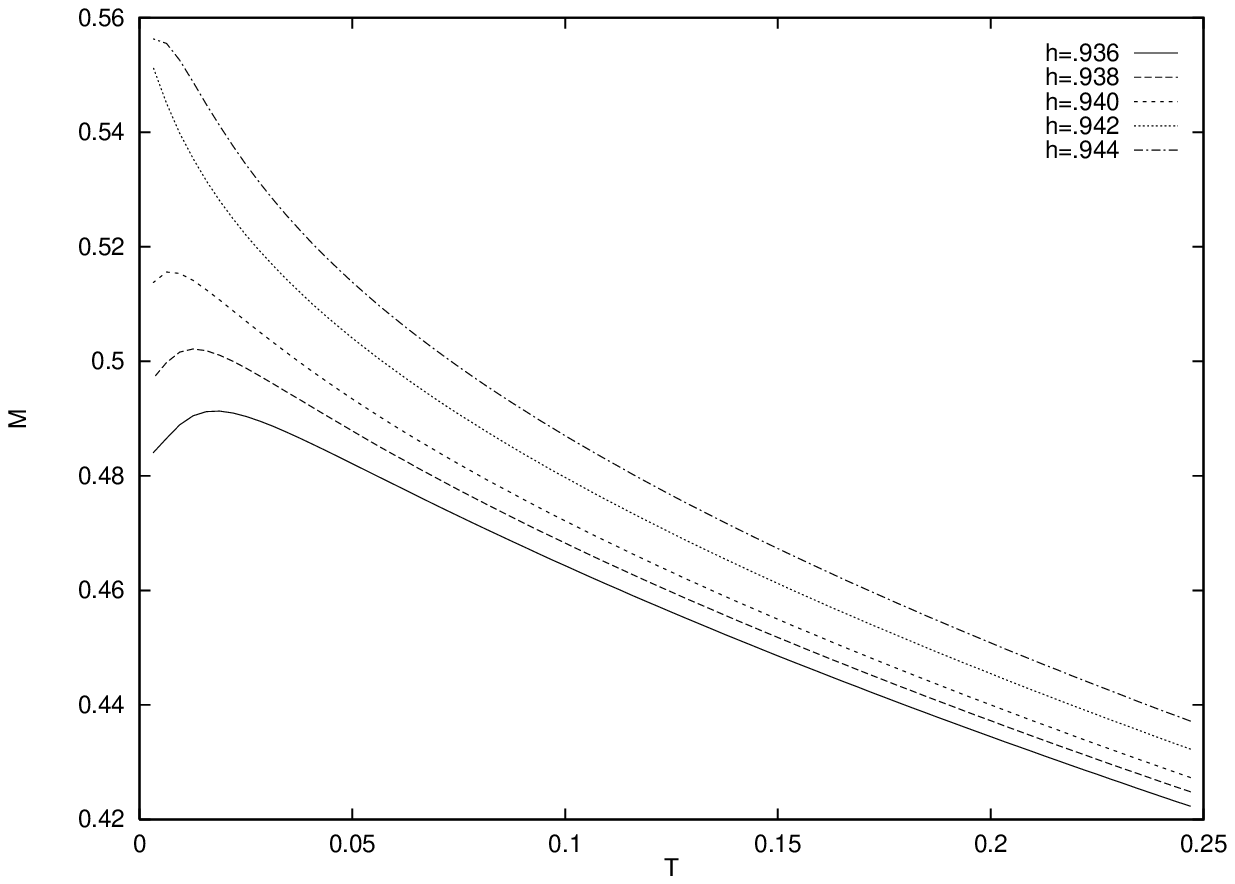}}
  \centerline{Fig.5b}
\end{figure}
\newpage
\begin{figure}
  \epsfxsize = 12 cm   
  \centerline{\epsfbox{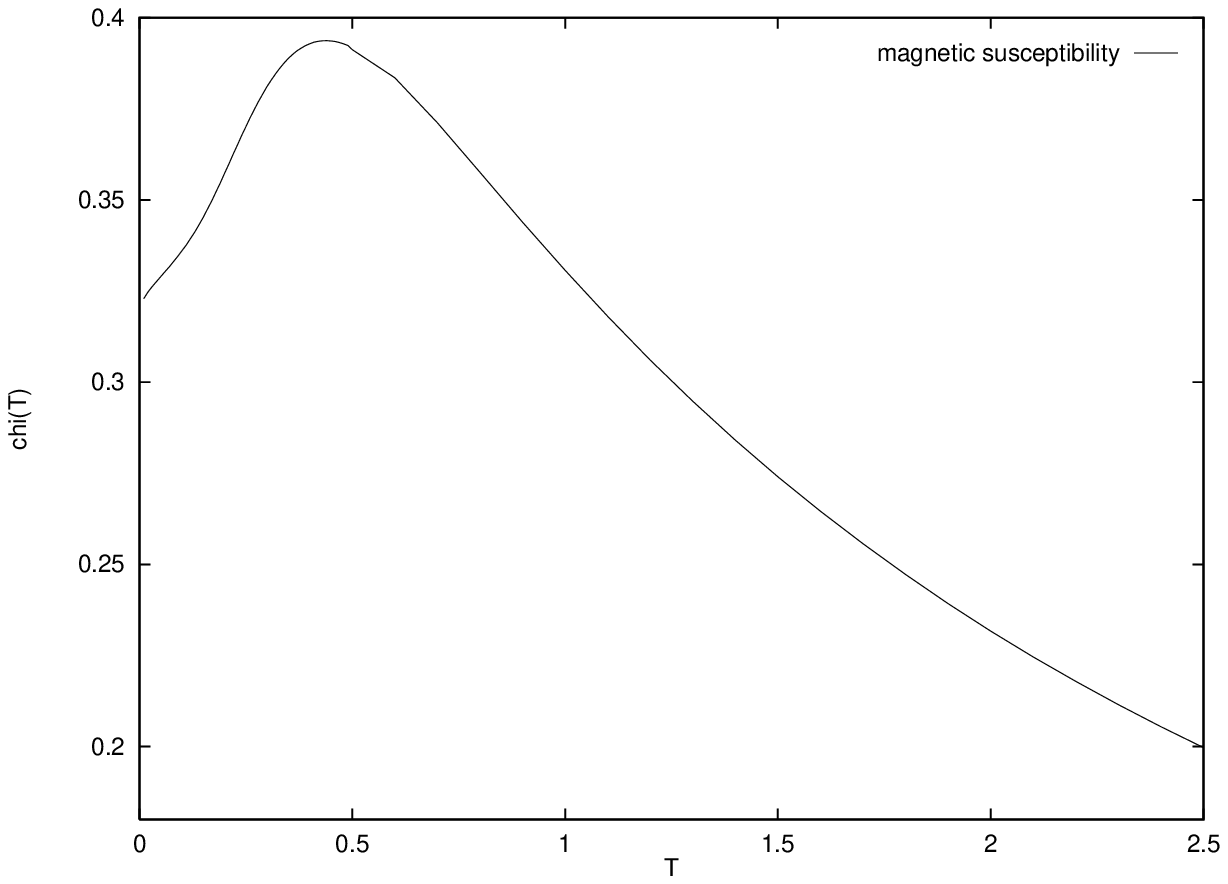}}
  \centerline{Fig.6}
\end{figure}
\end{document}